\documentclass[aps,pre,10pt,onecolumn,tightenlines,superscriptaddress,amsmath,amssymb,final,letterpaper,notitlepage,nofootinbib]{revtex4-1}
\usepackage{graphicx}
\usepackage{subfigure}
\usepackage{epstopdf}
\usepackage{times}
\usepackage{bm}
\usepackage[dvipsnames]{xcolor}
\usepackage{tcolorbox}
\usepackage{hyperref}

\hypersetup{bookmarks=true, unicode=false, pdftoolbar=true, pdfmenubar=true, pdffitwindow=false, pdfstartview={FitH}, pdfnewwindow=true, colorlinks=true, linkcolor=blue, citecolor=blue, filecolor=magenta, urlcolor=blue}
\graphicspath{{figs/}} 

\begin{document}
\title{Interacting contagions are indistinguishable from social reinforcement}

\author{{Laurent \surname{H\'ebert-Dufresne}}}
\affiliation{Vermont Complex Systems Center, University of Vermont, Burlington, VT 05405, USA}
\affiliation{Department of Computer Science, University of Vermont, Burlington, VT 05405, USA}
\affiliation{D\'epartement  de  physique,  de  g\'enie  physique et  d'optique, Universit\'e  Laval,  Qu\'ebec  (Qu\'ebec),  Canada  G1V  0A6}
\author{{Samuel V. \surname{Scarpino}}}
\affiliation{Network Science Institute, Northeastern University, Boston, MA 02115, USA}
\affiliation{Marine \& Environmental Sciences, Northeastern University, Boston, MA 02115, USA}
\affiliation{Physics, Northeastern University, Boston, MA 02115, USA}
\affiliation{Health Sciences, Northeastern University, Boston, MA 02115, USA}
\affiliation{ISI Foundation, 10126 Turin, Italy}
\author{{Jean-Gabriel \surname{Young}}}
\affiliation{Center for the Study of Complex Systems, University of Michigan, Ann Arbor, MI 48109, USA}
\begin{abstract}
From fake news to innovative technologies, many contagions spread via a process of social reinforcement, where multiple exposures are distinct from prolonged exposure to a single source. Contrarily, biological agents such as Ebola or measles are typically thought to spread as simple contagions.  Here, we demonstrate that interacting simple contagions are indistinguishable from complex contagions. In the social context, our results highlight the challenge of identifying and quantifying mechanisms, such as social reinforcement, in a world where an innumerable amount of ideas, memes and behaviors interact. In the biological context, this parallel allows the use of complex contagions to effectively quantify the non-trivial interactions of infectious diseases.
\end{abstract}
\maketitle

On September 27th 2016, the World Health Organization (WHO) declared that measles had been eliminated from the Americas~\cite{PAHO2106}. Less than two years later, an outbreak of the disease in Venezuela sparked an epidemic across South America, which is ongoing and has sickened tens-of-thousands~\cite{dabbagh2018progress, fraser2018measles, elidio2019measles}.  Concurrently, the number of measles cases has increased in all but one of the WHO regions~\cite{friedrich2019measles}, over 80,000 cases (with a hospitalisation rate $>$60\%) occurred in the European Union~\cite{thornton2019measles}, and the United States of America experienced 17 measles outbreaks~\cite{dabbagh2018progress, CDCMEAS2019}.  The majority of these cases occurred in unvaccinated individuals~\cite{majumder2015substandard, phadke2016association, melegaro2019measles}.  From collapsing public health infrastructure~\cite{paniz2019resurgence} and lack of access to vaccines~\cite{elidio2019measles} to non-medical exemptions, e.g., religious beliefs~\cite{salmon1999health, papachrisanthou2019resurgence}, and the spread of fraudulent science~\cite{mchale2016reasons, sansonetti2018measles, mavragani2018internet}, the precise reason individuals go unvaccinated are myriad; however, underlying all these mechanisms is the coupled transmission of two contagions, one biological and one---or more---social.  

Clearly, contagions never occur in a vacuum, instead pathogens and ideas interact with each other and with externalities such as host connectivity, behaviour, and mobility. Nevertheless, many biological contagions are still considered to be ``simple", where infectious individuals transmit to susceptible individuals independently of anything else occurring around the individuals, susceptible and/or infectious, in question \cite{anderson1992infectious}. In complex contagions, however, the spreading mechanism explicitly depends on the context of transmission events, usually via the neighbourhood of the susceptible individuals, such that pairwise information becomes insufficient to model the transmission process \cite{weng2013virality,monsted2017evidence}. For example, social reinforcement can lead to a transmission rate effectively proportional to the number of different infectious contacts to which a susceptible individual is exposed \cite{centola2010spread}. This mechanistic difference creates a false dichotomy, forcing us to choose the mechanism we think best describe the reality of a given contagion. In practice, the context of transmission events always matters.

When modeling a contagion, the choice of mechanism is critical because simple and complex contagions tend to induce substantially different dynamics and can lead to incompatible conclusions about intervention strategies or risk. An important difference is that complex contagions do not always feature a monotonous relation between the expected epidemic size and their average transmission rate, unlike simple contagions \cite{dodds2004universal}. Instead, microscopic variations in transmission rate can lead to macroscopic jumps in expected epidemic size. This effect, that small changes in transmission can lead to large differences in outbreak size, occurs because the population effectively builds up a latent epidemic potential where many individuals would infect their susceptible neighbours if only a few of them had one more infectious neighbour. Eventually, often due only to small variation in initial conditions or transmission rate, a macroscopic cascade of infections that releases this latent epidemic potential will occur. The importance of correlations between the states of neighbours also explains why complex contagions can benefit from network clustering (i.e. triangles), again unlike simple contagions \cite{osullivan2015mathematical}.

Multiple recent studies have shown that interacting simple contagions share many of the same defining features as complex contagions, such as their discontinuous phase transitions and how they benefit from network clustering \cite{hebert2015complex}. Additionally, seemingly different formulations of the complex contagion model, e.g., the voter model vs. a bursty threshold model, may belong to the same universality class as defined by the models' epidemic thresholds on power law networks with $\gamma<5/2$~\cite{cota2018robustness}. Consider for example influenza and Streptococcus pneumoniae, which can interact in different ways: An individual with a compromised immune system due to one infection might be more susceptible to the other, or an individual with both infections might exhibit heightened symptoms and increased transmission rates. A population can then build up a latent epidemic potential where many individuals would infect their susceptible neighbours if only a few of them were compromised by a second disease.  Therefore, it might not be surprising that interacting contagions can also exhibit macroscopic jumps in expected epidemic size under microscopic variation in the average transmission rates, and that interacting contagions can also benefit from network clustering.  Even for single pathogens spreading through a population, if different routes of transmission have drastically different transmission probabilities, e.g., HIV sexual transmission vs. needle sharing~\cite{strathdee2010epidemiology, volz2010epidemiological} or Zika sexual transmission from men vs. women~\cite{allard2017asymmetric}, they will spread more like complex contagions.

Here, we demonstrate that the correspondence between complex and interacting contagions runs far deeper. In Box 1, we present simple models of complex and interacting contagions in well-mixed populations and demonstrate that there exists an exact mapping between the two models, i.e. the two mechanisms are indistinguishable. Figure 1 summarizes the interesting dynamical features of these models, namely their potential for discontinuous jumps in expected epidemic size as well as regimes of faster than exponential spread. Figure 1 also highlights the mathematical mapping between the two dynamics.  Every curve of Fig. 1 is obtained by solving the differential equations for both interacting and complex contagions. The solutions are identical. Consequently, it follows that in any context where the assumption of a well-mixed population holds, complex and interacting contagion models are indistinguishable-- provided we are unaware of potential interactions among pathogens, and are not collecting the corresponding coinfection data. Given that both assumptions, well-mixed populations and data from a single pathogen, are often considered to hold in contained environments such as schools, workplaces, and homogenous social groups \cite{wearing2005appropriate}, our results demonstrate that, unless the process follows strictly simple dynamics, even perfect incidence data for a single contagion in these environments cannot be used to identify the true spreading mechanism.

\begin{tcolorbox}[title= Box 1: Well-mixed compartmental SIS models]

We use the simple Susceptible-Infectious-Susceptible (SIS) process to highlight the mathematical mapping between complex contagions and interacting simple contagions. In the SIS process, infectious individuals infect susceptible individuals, but also recover back to the susceptible state. A general complex contagion model can be followed using an ordinary differential equation (ODE) for the density $I(t)$ of infectious individuals at time $t$. Omitting the obvious temporal dependency, we write
\begin{equation}
\dot{I} = \beta\left(I\right) I\left(1-I\right) - \gamma I \label{eq:I}
\end{equation}
where $\gamma$ is the recovery rate and where $\beta\left(I\right)$ describes the transmission rate per contact given that there is a density $I(t)$ of infectious individuals in the population. For example, an increasing $\beta\left(I\right)$ can describe a social reinforcement mechanism.\\

We can write similar equations for two interacting simple contagions by tracking the density $[SS]$ of individuals which are susceptible to both contagions, the densities $[IS]$ and $[SI]$ that are infected by the first or second contagion only, and the density $[II]$ that are coinfected. We are mostly interested in the density of nodes $\tilde{I}=[IS] + [II]$ which are infected by the first contagion. We therefore follow
\begin{align}
\dot{[II]} = & [IS]\left(\beta_2[SI]+\alpha\beta_2[II]\right) +[SI]\left(\beta_1[IS]+\alpha\beta_1[II]\right)-(\gamma_1+\gamma_2)[II] \label{eq:II} \\
\dot{[IS]} = & [SS]\left(\beta_1[IS]+\alpha\beta_1[II]\right) + \gamma_2[II] -\gamma_1[IS] - [IS]\left(\beta_2[SI]+\alpha\beta_2[II]\right) \label{eq:IS} \; .
\end{align}

Comparing the sum of Eqs.~(\ref{eq:II}) and (\ref{eq:IS}) with Eq.~(\ref{eq:I}) shows that the two models are equivalent under the mapping
\begin{equation}
\beta(I) \equiv \beta(\tilde{I}) = \dfrac{\beta_1[IS] + \alpha\beta_1[II]}{[IS]+[II]}
\end{equation}
which is a valid mapping from interacting contagions to a complex contagion, for every monotonous form of $[IS]+[II]$ (the vast majority of possible SIS curves). A similar mapping can be obtained for Susceptible-Infectious-Recovered dynamics using the monotonicity of the recovered curve assuming recovered individuals are now immune.

\end{tcolorbox}

\begin{figure}[t!!!!!!!]
\centering
\includegraphics[width=0.99\linewidth]{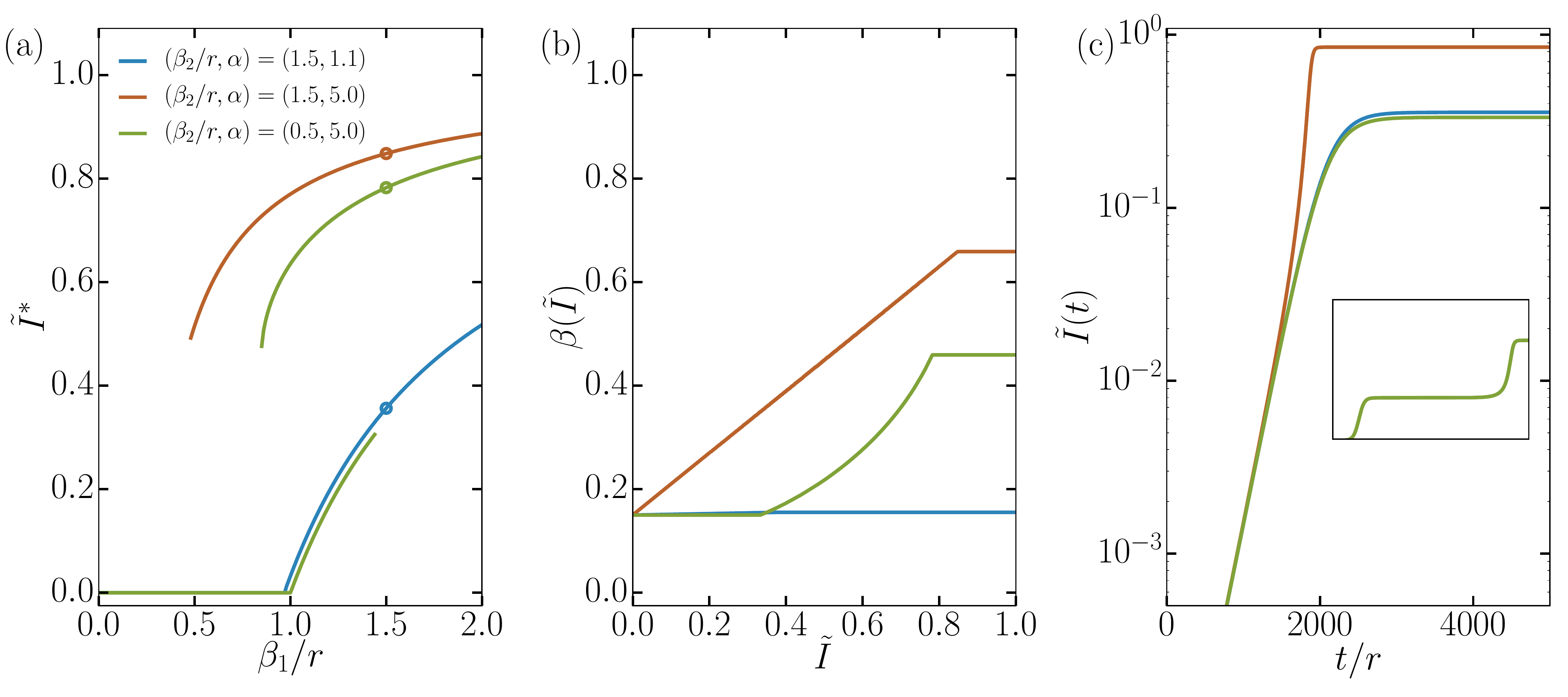}
\caption{\label{F1}\textbf{Analytical solutions of \textit{both} SIS ODE systems from Box 1: Interacting simple contagions and complex contagions.} In panel \textit{a)} we present the expected final epidemic size, which can undergo a continuous, hybrid, or discontinuous transition at varied epidemic thresholds. We highlight three points whose time evolution is explored in panel \textit{b)} and \textit{c)} In panel \textit{b)} we present the effective transmission rate per infectious contact, which are monotonously increasing in all cases. In the continuous transition regime (blue curve), $\beta(\tilde{I})$ is almost but not quite constant at $\beta_1$. In the hybrid transition regime (green curve), $\beta_1$ is flat until $\tilde{I}$ reaches the expected epidemic size $\tilde{I}^*$ of its second epidemic transition, at which points the synergistic interactions start. In the discontinuous transition regime (orange curve), $\beta(\tilde{I})$ is a linear function of $\tilde{I}$. Panel \textit{c)} illustrates the varied time evolution of $\tilde{I}(t)$ through time. In the continuous regime (blue curve), we see an exponential growth with saturation at equilibrium. In the discontinuous regime (orange curve), we see an exponential growth followed by a regime of rapid acceleration. In the hybrid regime (green curve), we see both behaviours in an evolution with two plateaus; mostly visible on the inset which shows the longer, complete time evolution. All results were obtained both by integrating Eqs.~\eqref{eq:II}--\eqref{eq:IS}, and by using the obtained $\beta(\tilde{I})$ curve as an input to Eq.~\eqref{eq:I}. We found that the two approaches, interacting simple contagions or complex contagions, give the exact same results in all regimes.}
\end{figure}

The situation is different in heterogeneous environment, where contagions follow some underlying contact network. In this context, a contagion process is not fully described by simply following the number of infections in time, and the identity and contacts of infected individuals matters for the dynamics. In Ref.~\cite{weng2013virality}, the authors study memes by quantifying their spread within and across communities.  Their approach is based on the fact that for complex contagions, unlike simple contagions, ``the spread within highly clustered communities is enhanced, while diffusion across communities is hampered.'' They identify a few statistics that can help distinguish simple and complex contagions such as the number of early adopters of memes (number of contagious nodes at early times) and the state of their neighbours. The logic is that if contagions benefit from network clustering, states of neighbours should be more correlated than expected from the network structure alone.

Unfortunately, interacting contagions also benefit from network clustering \cite{hebert2015complex}. Using simulations--where the underlying network structure is known--we find that looking for state correlations between neighbours on a contact networks can indeed distinguish interacting and complex contagions from simple contagions as claimed in Ref.~\cite{weng2013virality}, but unfortunately can not distinguish complex and interacting contagions from one another. Our experiments consist of simple, interacting and complex contagions simulated on both regular random contact networks (20 contacts per node) and equivalent but clustered networks where cliques of size 12 are formed around every node for a clustering coefficient $C=0.29$. Different interaction mechanisms can be considered in the case of interacting contagions in simulations, but most importantly we increase the transmission rates of both contagions by a factor of 7 when they appear on the same contact. This interaction parameter might appear high but is actually far from, for example, the interaction factor between influenza and pneumococcal pneumonia which can be up to 100-fold ~\cite{shrestha2013identifying}.

\begin{figure}
\centering
\includegraphics[width=0.3\linewidth]{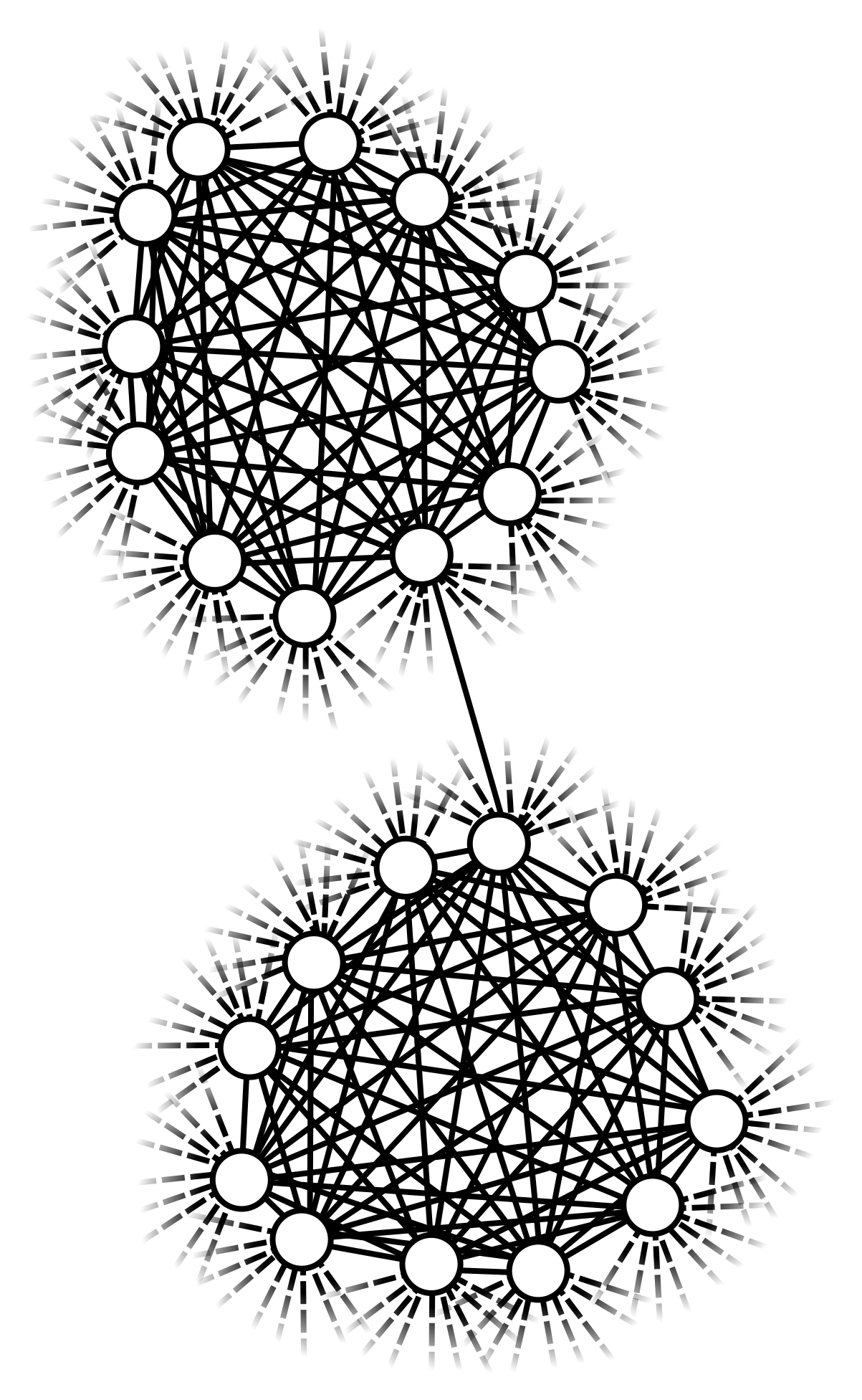}
\includegraphics[width=0.6\linewidth]{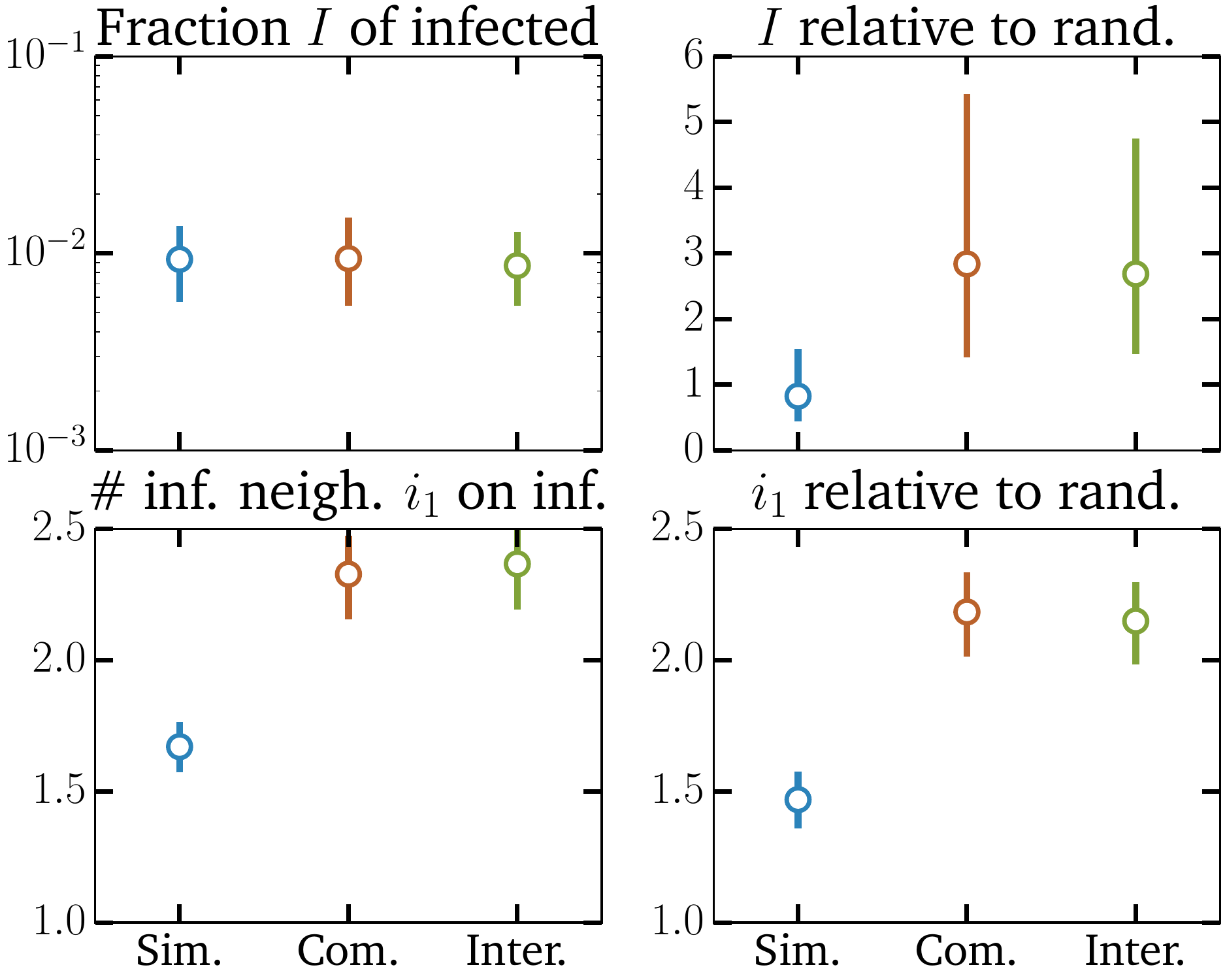}
\caption{\textbf{Statistics of SIS simulations of simple, interacting and complex contagions.} (left) The contact network structure used in the simulations. All $10^4$ nodes belong to a single clique of size 12, and also have an additional 9 random neighbors. (right) Observed phenomenology: fraction of infected nodes 5000 time steps and average number of infectious neighbors upon infection of a node, as well as a comparison of the same features versus expectations on an equivalent random network (same degree distribution, but all edges are uniformly randomized). We show the median value of all statistics as well as their 50\% confidence interval. Parameters were chosen such that the fraction of infected nodes are similar after the 5000 time steps. The simple contagion uses a transmission probability of $8\cdot 10^{-5}$ and a recovery probability of $10^{-3}$. The interacting simple contagions use a transmission probability of $12\cdot 10^{-6}$ and a recovery probability of $10^{-3}$, but the transmission probability increases by a factor of 15 if the other disease is present on the contact (in either the susceptible or infectious node) and the recovery probability decreases by a factor of 2 and 3 for the first and second disease if a node is infected by both. The statistics shown are for the first disease only. Finally, the complex contagion uses transmission and recovery probability that depend on the number $k$ of infectious neighbor. The transmission probability follows a sigmoid $p_0 + (3/4)p_0/(1+\textrm{exp}(-2(k-2))$ with $p_0 = 5\cdot10^{-5}$ whereas the recovery probability simply falls as $10^{3}/k$.}
\label{Fstats}
\end{figure}

The details of the experimental set-up are further discussed in Supplementary Material, but the results are straightforward (Fig.~\ref{Fstats}): When parametrized correctly such that all 3 contagion models can reach the same level of prevalence after some desired time period, we find that interacting and complex contagions can be easily distinguished from simple contagions, but not from one another.
This shows that not only do non-simple contagions benefit from clustering in network structure, but that network structure also leads to similar increases in statistical correlations between cases.

Altogether, the results from Box 1, Fig.~\ref{F1} and Fig.~\ref{Fstats} show that complex contagions and interacting simple contagions possess the same dynamical and statistical features, even if the mechanisms behind both models are completely different. Since we know real-world pathogens can interact with unknown numbers of other pathogens--or different strains/serotypes of the same pathogen--the logical conclusion is that we ought to model real contagions as complex contagions. Critically, these contagions need not all be biological and pathogens can interact with, for example, the spread of a behaviour. Additionally, simple contagions dynamics are a subset of the complex contagion model, meaning that if the pathogen does not interact with other contagions, the model can recapitulate its dynamics.  Additionally, by utilizing the complex contagion model, we avoid the need to procure exponentially more data and estimate exponentially more model parameters as the number of interacting contagions increases; instead we can use a general parametrization to find a complex transmission function $\beta(I)$ governing the effective infection rate for every pathogen of interest.

A similar conclusion was recently reached in Ref.~\cite{liu2018measurability} where a transmission rate that varies in time was used to follow an outbreak through a population.  Indeed, utilizing a time-varying transmission rate can also be used to mimic the impact of heterogeneous contact networks where the contagion time series might behave differently when moving from one community (e.g. a large dense school) to another (e.g. a smaller workplace). We thus develop a framework to infer a function $\beta(I)$ that reproduces some contagion time series from a complex contagion Markov model of Susceptible-Infectious-Recovered dynamics \cite{anderson1992infectious}. Essentially, this is the same well-mixed model as shown in Box 1 but where individuals who recover are now immune to the contagion instead of returning to the susceptible state. Because most real contagion data tend to be non-monotonous, i.e., they have a growth period, a peak, and a period of decay, the SIR model is likely more appropriate for empirical data. In this framework, we inverse the differential equations to infer a $\beta(I)$ function from data by assuming some Gaussian noise around the deterministic $I(t)$ equation and using a parametrization of $\beta(I)$ in terms of Bernstein polynomials.

We detail the inference procedure in our Supplementary Material. Importantly, our procedure focuses on $\beta(I)$ instead of an instantaneous transmission rate $\beta(t)$ as used in Ref.~\cite{liu2018measurability} for two reasons. First, using a $\beta(I)$ instead of an instantaneous $\beta(t)$ avoids overfitting to noise in the instantaneous growth rate of a time series $I(t)$, since we combine the information contained in both the rise and fall of a contagion. Second, using $\beta(I)$ maps real data to well-studied Markovian (i.e. memory-less) complex contagions and therefore also provides an easy way to compare contagions independent of time and initial conditions. 

\begin{figure}[t!!!!!!!]
\centering
\includegraphics[width=1\linewidth]{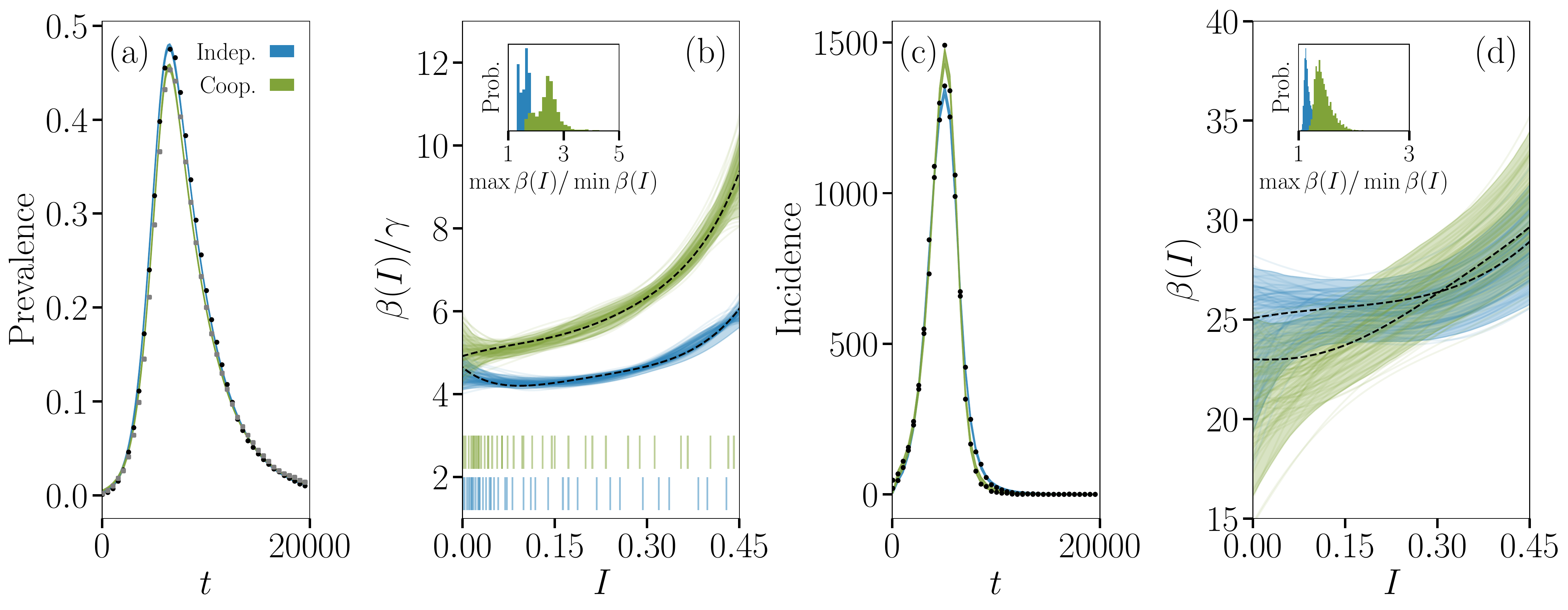}
\caption{\label{F3}\textbf{Signature of non-interacting and interacting SIR epidemics.} 
We simulate the Susceptible-Infectious-Recovered (SIR) process in the case of non-interacting and interacting simple contagions on clustered contact networks where each of 10,000 individuals belongs to 2 cliques of size 5. 
\textbf{(a)} Dynamics of the density of infected individuals for epidemics taking place on the same networks, and \textbf{(c)} corresponding incidence time series. 
Parameters were chosen to produce roughly similar time series.
In one case (blue data), two diseases with transmission rate $3/10$ and recovery rate $1/3$ spread without interaction, and we show the evolution of only one of them.
In the other case (green data), two diseases with transmission rate $1/10$ and recovery rate $1$ interact positively by increasing their transmission rate by a factor of 7 whenever the other disease occurs on the same contact and decreasing their recovery rate by the same factor when an individual is co-infected.
We again show the evolution of only one of the two diseases.
The colored curves are Bayesian fits of a continuous model of complex contagion in a well-mixed population  \cite{carpenter2017stan, jgyou_code}; the simulated time series are shown with small closed symbols.
\textbf{(b)} Inferred complex contagion function $\beta(I)/\gamma$ for the two series from prevalence data, with the density of observation illustrated with bar plots below.
\textbf{(d)} Same inference, but from incidence data.
The maximum a posteriori fit is shown with a dotted line, alonside with the 5\textsuperscript{th} to 95\textsuperscript{th} percentiles (shaded region), and 100 randomly selected posterior samples (transparent lines) that highlight the samples-to-samples variability. The insets describe how close the inferred transmission functions are to a flat transmission rate by plotting the distribution of the ratio of the maximal to minimal values in the posterior samples. In both cases, the transmission functions inferred on the time series of interacting contagions are significantly different from the transmission functions inferred on non-interacting contagion data.
}
\end{figure}

We test this inference procedure on simulations of both interacting and non-interacting simple contagions in Fig.~\ref{F3}. We selected parameter values to produce two very similar time series to test our framework on a hard instance of the contagion inference problem. While the only noise in the data is that of the regular stochasticity of epidemic simulations, we used a relative low sampling rate and used a single realization of both interacting and non-interacting contagion process. Even in this context we find that the strength of the interaction is reflected in how much the inferred $\beta(I)$ function deviates from a simple constant transmission rate, as measured by the ratio of the maximal to minimal values in the inferred $\beta(I)$ function. Our results therefore showcase how a robust framework of complex contagion can be used to quantify interactions between contagions.

In summary, interacting simple contagions and complex contagions are mathematically equivalent if we assume well-mixed populations. And, even if we know the underlying contact network, previously proposed statistics--such as the number of infectious neighbours upon infection or recovery--can only distinguish complex contagions or interacting contagions from simple contagions, but not from one another. Unfortunately, this implies that phenomenology and model fitting cannot distinguish or quantify spreading mechanisms if we are not fully aware of all possible interactions and coinfections.  Indeed, one consequence of our results is that measurements of complex spreading mechanisms, such as social reinforcement, are practically impossible.  Observations of the spread of an individual meme in a real social system will always be confounded by how an innumerable number of ideas, memes and behaviours might be interacting in the system unbeknownst to the modellers. Even worse, in practice we rarely have a perfect knowledge of the underlying contact network, such that the variations in transmission rate due to interactions are also combined with variations due to the unknown contact structure \cite{liu2018measurability}.

We therefore suggest embracing the reality that complex contagions can be useful even in context where we know the underlying mechanisms are wrong (i.e. can be falsified in experiments).  In fact, we claim that complex contagions are a general \textit{effective} framework for contagions of all nature.  Even when the idea of social reinforcement or non-linear infection rates have no mechanistic basis, they might be statistically justified because they can capture, infer, and quantify unknown interactions.

\subsection*{Acknowledgement}
L.H.D. acknowledges support from the National Science Foundations Grant No. NSF grant DMS-1829826 and the National Institutes of Health 1P20 GM125498-01 Centers of Biomedical Research Excellence Award. S.V.S. acknowledges support from startup funds provided by Northeastern University. J.G.Y. is supported by a James S. McDonnell Postdoctoral Fellowship. The authors also thank Antoine Allard, Benjamin M. Althouse, Mark Newman, George Cantwell, and Alec Kirkley for insightful discussions as well as John Burkardt for sharing his Bernstein polynomial implementation.
\subsection*{Data availability}
All data and software available from \cite{jgyou_code}.
\bibliographystyle{apsrev}

\end{document}